\documentstyle[a4,12pt,axodraw]{article}
\catcode `@=11 \@addtoreset{equation}{section} \catcode `@=12
\begin{document}
\def\be{\begin{equation}}
\def\ee{\end{equation}}
\def\ba{\begin{array}{l}}
\def\ea{\end{array}}
\def\bea{\begin{eqnarray}}
\def\eea{\end{eqnarray}}
\def\eq#1{(\ref{#1})}
\def\fig#1{Fig \ref{#1}} 
\def\wgnc{\bar{\wedge}}
\def\del{\partial}
\def\der{\overline \del}
\def\wg{\wedge}
\def\bull{$\bullet$}
\def\gap{\vspace{10ex}}
\def\tgap{\vspace{3ex}}
\def\sgap{\vspace{5ex}}
\def\lgap{\vspace{20ex}}
\def\half{\frac{1}{2}}
\def\pto{\vfill\eject}
\def\gst{g_{\rm st}}
\def\tC{{\widetilde C}}
\def\z{{\bar z}}
\def\o{{\cal O}}
\def\J{{\cal J}}
\def\S{{\cal S}}
\def\X{{\cal X}}
\def\N{{\cal N}}
\def\A{{\cal A}}
\def\H{{\cal H}}
\def\D{{\tilde D}}
\def\d{{\cal D}}
\def\re#1{{\bf #1}}
\def\nn{\nonumber}
\def\nl{\hfill\break}
\def\ni{\noindent}
\def\bibi{\bibitem}
\def\c#1{{\hat{#1}}}
\def\eps{{\epsilon}}
\pretolerance=1000000
\begin{flushright}
KEK/TH/756\\
April 2001\\
\end{flushright}
\begin{center}
\vspace{2 ex}
{\large\bf Loop Equation and Wilson line Correlators in Non-commutative 
Gauge Theories}
\\
\vspace{3 ex}
Avinash Dhar $^{1*}$ and Yoshihisa Kitazawa $^{\dagger}$ \\
~\\
{\sl Laboratory for Particle and Nuclear Physics,}\\
{\sl High Energy Accelerator Research Organization (KEK),}\\
{\sl Tsukuba, Ibaraki 305-0801, JAPAN.} \\

\vspace{10 ex}
\pretolerance=1000000
\bf ABSTRACT\\
\end{center}
\vspace{1 ex} 

We investigate Schwinger-Dyson equations for correlators of Wilson
line operators in non-commutative gauge theories. We point out that,
unlike what happens for closed Wilson loops, the joining term survives
in the planar equations. This fact may be used to relate the
correlator of an arbitrary number of Wilson lines eventually to a set
of {\it closed} Wilson loops, obtained by joining the individual
Wilson lines together by a series of well-defined cutting and joining
manipulations. For closed loops, we find that the non-planar
contributions do not have a smooth limit in the limit of vanishing
non-commutativity and hence the equations do not reduce to their
commutative counterparts. We use the Schwinger-Dyson equations to
derive loop equations for the correlators of Wilson observables. In
the planar limit, this gives us a {\it new} loop equation which
relates the correlators of Wilson lines to the expectation values of
closed Wilson loops. We discuss perturbative verification of the loop
equation for the $2$-point function in some detail. We also suggest a
possible connection between Wilson line based on an arbitrary contour
and the string field of closed string.

\vfill
\hrule
\vspace{0.5 ex}
\leftline{$^1$ On leave from Dept of Theoretical Phys, Tata Institute, Mumbai
400005, INDIA.}  
\leftline{$^*$ adhar@post.kek.jp}
\leftline{$^\dagger$ kitazawa@post.kek.jp}

\clearpage

\vspace{8ex}

\section{Introduction} 

Non-commutative gauge theories are realized on branes in the zero slope
limit in the presence of a large NS-NS B-field
\cite{CDS,DH,AAJ,CK,C2,CH,VS,SW1}. Recently these theories have
attracted a lot of attention. Various aspects of these theories have
been studied in
\cite{MR,HI,NS,AIIKKT,L,C1,I,IIKK1,SW2,DMWY,DRT,BM,AMNS,MRS,AD,IIKK2,DG,
GW,F,JMW,HKLM,DMR,W,GMS,S,KRS,MW,DK1}.

In ordinary gauge theories, a generic gauge-invariant observable is
provided by an arbitrary closed Wilson loop. Non-commutative gauge
theories have more general gauge-invariant observables, defined on
{\it open} contours. Such gauge-invariant observables in
non-commutative gauge theories were constructed in \cite{IIKK1}.
Different aspects of these were studied in
\cite{AD,IIKK2,RU,DR,GHI,DW,DK2,HL,DT,OO,LM,O}. Roughly speaking, these
gauge-invariant observables can be written as Fourier transforms of
open Wilson lines. In the operator formalism they are given by the
following expression
\bea 
W_C[y]={\rm Tr}\bigg(Pexp\{i\int_C d\sigma \ \del_\sigma y_\mu(\sigma)
A_\mu(\hat x+y(\sigma))\} \ e^{ik.\hat x} \bigg),
\label{oneone}
\eea
where
\bea
[{\hat x}_\mu, {\hat x}_\nu]=i\theta_{\mu\nu},
\label{onetwo}
\eea
and the trace in (\ref{oneone}) is over both the gauge group, taken
here to be $U(N)$, as well as the operator Hilbert space. These open
Wilson lines are gauge-invariant in non-commutative gauge theories,
unlike in ordinary gauge theories, provided the momentum $k^\mu$
associated with the Wilson line is fixed in terms of the straight line
joining the end points of the path $C$, given by $y^\mu(\sigma)$ where
$0 \leq \sigma \leq 1$, by the relation 
\bea
y_\mu(1)-y_\mu(0)=\theta_{\mu\nu}k_\nu.
\label{onethree}
\eea
The path $C$ is otherwise completely arbitrary. When $k$ vanishes, the two
ends of the path $C$ must meet and we have a closed Wilson loop.

In an earlier work \cite{DK2}, based on a perturbative analysis of
correlation functions of straight Wilson lines with generic momenta,
we had suggested that at large momenta the Wilson lines are bound into
the set of closed Wilson loops that can be formed by joining the
Wilson lines together in all possible different ways. In the present
work we will establish a more general connection between correlators
of Wilson lines and expectation values of Wilson loops in a
non-perturbative setting, for arbitrary Wilson lines. In this generic
case, however, the closed Wilson loops to which the Wilson lines are
related are not formed by simply joining the Wilson lines together,
but by more complicated cutting and joining manipulations. Also, the
statement is valid for arbitrary momenta, not necessarily large,
carried by the Wilson lines.  We will use the framework of
Schwinger-Dyson equations and the closely related loop equations in
the planar limit. In the context of non-commutative gauge theories
similar equations have been studied earlier in \cite{FKKT,DW2,AZD}.

This paper is organized as follows. In the next section we summarize
some aspects of operator formulation of non-commutative gauge
theories, which is used throughout this paper, and in particular list
some useful identities. In Sec.3 we derive Schwinger-Dyson equations
for the correlators of open and closed Wilson observables and discuss
these at finite $N$ as well as in the planar limit. As in commutative
gauge theories, the splitting term disappears from the planar
equations.  However, unlike in the case of closed Wilson loops, the
joining term survives in the planar equations for open Wilson lines.
This has the consequence of eventually relating them to closed Wilson
loops. We also find that at finite $N$, the splitting term does not
reduce to the ordinary gauge theory result in the limit in which the
non-commutativity is removed. We trace this result to the UV-IR mixing
in non-commutative gauge theories.  In Sec.4 we use the results of
Sec.3 to write down loop equations for the correlators of open and
closed Wilson observables.  We consider the loop equation for the
$2$-point function of the open Wilson lines in the planar limit and
discuss the verification of this equation in 't Hooft perturbation
theory in some detail. Sec.5 contains a discussion of a possible
connection between a Wilson line based on an arbitrary contour and
string field for closed string and the non-perturbative meaning of the
new loop equations derived here.  In the Appendix, we give details of
the perturbative calculations.

\section{Non-commutative gauge theories - operator formulation} 

We will be working in $4$-dimensional Euclidean space with a generic
non-commutativity parameter in (\ref{onetwo}). The non-commutative gauge
theory action that we will consider is 
\bea 
S={1 \over 4g^2}{\rm Tr}(F_{\mu\nu}(\hat x))^2 + \cdots
\label{twoone}
\eea
where
$$F_{\mu\nu}(\hat x)=\hat\del_\mu A_\nu(\hat x)-\hat\del_\nu
A_\mu(\hat x) +i[A_\mu(\hat x),A_\nu(\hat x)].$$ The dots stand for
possible bosonic and fermionic matter coupled to the gauge field and
the trace ${\rm Tr} = {\rm tr}_{U(N)} \ {\rm tr}_{\cal H}$ is over the
gauge group $U(N)$
as well as the operator Hilbert space $\cal H$. To define this latter
trace more precisely, let us assume that $\theta_{\mu\nu}$ has the
canonical form, 
\bea
\theta_{01}=-\theta_{10}=\theta_a, \ 
\theta_{23}=-\theta_{32}=\theta_b,
\label{twoonea}
\eea
with all other components vanishing, and let us define the operators
\bea
a={{\hat x}_0 +{\hat x}_1 \over \sqrt{2\theta_a}}, \ b={{\hat x}_2
+{\hat x}_3 \over \sqrt{2\theta_b}},
\label{twooneb}
\eea
which satisfy the standard harmonic oscillator algebra,
\bea
[a,a^\dagger]=1, \ [b,b^\dagger]=1.
\label{twoonec}
\eea
The operator Hilbert space trace is then defined by 
\bea
{\rm tr}_{\cal H} \hat O(\hat x)=(2\pi)^2 \theta_a \theta_b \sum_{n_a,n_b} 
<n_a,n_b|\hat O(\hat x)|n_a,n_b>.
\label{twooned}
\eea
Note that with this definition of the operator Hilbert space trace,
the coupling constant $g$ appearing in the action (\ref{twoone}) is
dimensionless.

We use the standard Weyl operator ordering,
\bea
\hat O(\hat x)= \int d^4y \ O(y) \ \delta^{(4)}(\hat x - y)
\label{twotwo} 
\eea
where the {\it operator} delta-function is defined in terms of the 
Heisenberg group elements by
\bea
\delta^{(4)}(\hat x - y)=\int {d^4k \over (2\pi)^4} \ e^{-ik.y} \ 
e^{ik.\hat x}.
\label{twothree}
\eea
The use of this operator delta-function simplifies many calculation
because it shares some properties of the usual delta-function. For
example, (\ref{twotwo}) and 
\bea
{\rm tr}_{\cal H}\delta^{(4)}(\hat x - y)=1. 
\label{twofour}
\eea
There are, of course, differences as in the following identity which
encodes the star product:
\bea
\delta^{(4)}(\hat x - y) \ \delta^{(4)}(\hat x - z) 
=e^{-{i \over 2} \del_+ \theta \del_-} \delta^{(4)}(\hat x-y_+)
\delta^{(4)}(y_-).
\label{twofive}
\eea
where $y_+={y+z \over 2}$ and $y_-=y-z$.
Below we give two identities involving these operator delta-functions
which will be used in deriving the Schwinger-Dyson equations in the
next section. The first one ``joins'' together two operators which
appear inside two different traces,
\bea
\int d^4z \ \sum_a \ {\rm Tr}[\hat O_1(\hat x) t^a \delta^{(4)}(\hat x - z)]
{\rm Tr}[\hat O_2(\hat x) t^a \delta^{(4)}(\hat x - z)]
= {\rm Tr}[\hat O_1(\hat x) \hat O_2(\hat x)],
\label{twosix}
\eea
and the second one ``splits'' two operators which are inside the same trace,
\bea
\int d^4z \ \sum_a \ {\rm Tr}[\hat O_1(\hat x) t^a \delta^{(4)}(\hat x - z)
\hat O_2(\hat x) t^a \delta^{(4)}(\hat x - z)]
= {1 \over (2\pi)^4 {\rm det}\theta} {\rm Tr}[\hat O_1(\hat x)] 
{\rm Tr}[\hat O_2(\hat x)]. \nonumber \\
\label{twoseven}
\eea
Here the $t^a$'s are the generators for the gauge group, which we have
taken to be U(N), with the normalization dictated by the completeness
condition
\bea
\sum_a t^a_{ij}t^a_{kl}=\delta_{il}\delta_{jk}
\label{twosevena}
\eea

\subsection{Wilson observables and cyclic symmetry}

The generic gauge-invariant Wilson observable is given in
(\ref{oneone}). We will also need the Wilson operator
\bea 
\hat W_C[y]_{0s}=Pexp\{i\int_C d\sigma \ \del_\sigma y_\mu(\sigma)
A_\mu(\hat x+y(\sigma)-y(0))\} \ e^{ik_s.\hat x}.
\label{twoeight}
\eea
Here the subscripts `$0s$' indicate that the path-ordered phase factor
runs from $\sigma=0$ to $\sigma=s$ along the curve $C$, and
$y(s)-y(0)=\theta k_s$. The Wilson operator $\hat W_C[y]_{s1}$, which
runs from $\sigma=s$ to $\sigma=1$, is defined similarly: 
\bea
\hat W_C[y]_{s1}=Pexp\{i\int_C d\sigma \ \del_\sigma y_\mu(\sigma)
A_\mu(\hat x+y(\sigma)-y(s))\} \ e^{i\tilde k_s.\hat x} 
\label{twonine}
\eea
where $y(1)-y(s)=\theta \tilde k_s$. These operators are related to the 
Wilson observable as follows:
\bea
{\rm Tr}(\hat W_C[y]_{01})=W_C[y] \ e^{ik.y(0)}.
\label{twoten}
\eea 

The Wilson observable $W_C[y]$ possesses a ``cyclic symmetry'' because of
the trace over both the gauge group and the operator Hilbert space. To
arrive at a mathematical expression of this symmetry, we note that
\bea
\hat W_C[y]_{01}=e^{{i \over 2}k_s \theta k} \ \hat W_C[y]_{0s}
\hat W_C[y]_{s1},
\label{twoeleven}
\eea
and, \footnote{In the following $\hat W_C[y]_{01}$, which goes over the 
full parametric range, will be denoted by $\hat W_C[y]$ for notational 
convenience.} therefore,
\bea
W_C[y] &=& e^{-ik.y(0)} \ e^{{i \over 2}k_s \theta k} \ 
{\rm Tr}(\hat W_C[y]_{0s} \hat W_C[y]_{s1}) \nonumber \\
&=& e^{-ik.y(s)} \ {\rm Tr}(\hat W_{C_s}[y_s]) \nonumber \\ 
&\equiv&  W_{C_s}[y_s],
\label{twotwelve}
\eea
where in the second step we have used the cyclic property of the trace
and recombined the two operators in the opposite order. The contour
$C_s$ is given by
\bea
y_s(\sigma) &=& y(\sigma + s), \ \ \ 0 \leq \sigma \leq (1-s) 
\nonumber \\
&=& y(\sigma-1+s)+y(1)-y(0), \ \ (1-s)\leq \sigma \leq 1.
\label{twothirteen}
\eea
It is obtained from the curve $C$ by cutting it at a point $\sigma=s$
and rejoining the two pieces in the opposite order, as shown in Fig.
1. 

\begin{center} 
\begin{picture}(400,150)(0,0)
\ArrowArc(150,75)(50,90,270)
\ArrowArc(300,125)(50,180,270)
\ArrowArc(300,25)(50,90,180)
\LongArrow(160,75)(230,75)
\Text(158,75)[r]{$C$}
\Text(150,129)[lb]{$\sigma=0$}
\Text(98,75)[rt]{$s$}
\Text(150,25)[lt]{$\sigma=1$}
\Text(235,75)[l]{$C_s$}
\Text(250,129)[lb]{$\sigma=0$}
\Text(302,75)[l]{$(1-s)$}
\Text(250,25)[lt]{$\sigma=1$}
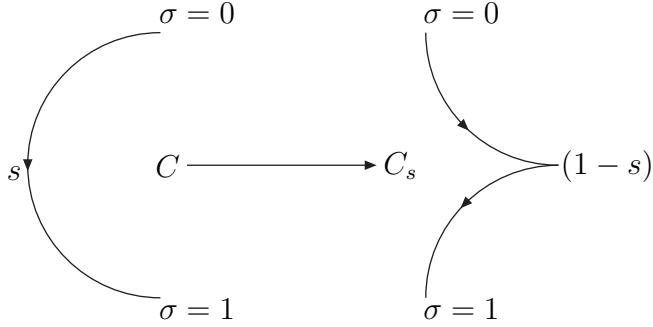
\end{picture} \\ {\sl Fig. 1: Cyclic symmetry of Wilson line}
\end{center}

If the original curve is a straight line, then the transformed curve
is also a straight line shifted by an amount $s(\theta k)$. It is easy
to see more directly that such shifts are a symmetry of the straight
Wilson line. This symmetry was used very effectively in \cite{DK2} to
simplify perturbative calculations. More generally, the cyclic
symmetry relates Wilson observables defined on contours that are
nontrivially different. As we shall see later, the quantity
\bea
V^{(k)}_{C\mu}[y]=i\int_C dy_{\mu}(\sigma) \ e^{-ik.y(\sigma)} 
\label{twofourteen}
\eea 
frequently appears in perturbative calculations of multipoint Wilson
line correlation functions. It is easy to see that in fact this
quantity is invariant under the cyclic symmetry (\ref{twothirteen}),
and that may be reason for its appearance. It is also noteworthy that
the above quantity is very similar to the vector vertex operator of
open string theory. It would be interesting to have a better
understanding of these connections and the implications of the cyclic
symmetry.

\section{Schwinger-Dyson equation}

In this section we will first derive the Schwinger-Dyson equation for
multipoint correlators of Wilson observables and then analyse it at
finite $N$ as well as in the planar limit. The Schwinger-Dyson
equation follows from the standard functional integral identity
\bea
0=\int[{\cal D}A_\mu^b(x)] \int d^4z \sum_a {\delta \over 
\delta A_\mu^a(z)} \bigg[e^{-S} \ {\rm Tr}(\hat W_{C_1}[y_1]t^a
\delta^{(4)}(\hat x-z))
{\rm Tr}(\hat W_{C_2}[y_2]) \cdots {\rm Tr}(\hat W_{C_n}[y_n])\bigg].
\label{threeone}
\eea
Using 
\bea
{\delta \over \delta A_\mu^a(z)} \hat W_C[y]=i \int_C dy_\mu(s) \
e^{{i \over 2}k_s \theta k} \ \hat W_C[y]_{0s}(t^a\delta^{(4)}(\hat x-z))
\hat W_C[y]_{s1}
\label{threetwo}
\eea
and the joining and splitting identities, (\ref{twosix}) and (\ref{twoseven}),
we get
\bea
&&{1 \over g^2} <{\rm Tr}(\hat W_{C_1}[y_1]\hat D_\nu\hat F_{\mu\nu}(\hat x))
{\rm Tr}(\hat W_{C_2}[y_2]) \cdots {\rm Tr}(\hat W_{C_n}[y_n])>  \nonumber \\
&=& i\sum^n_{l=2}\int_{C_l}dy_{l\mu}(s) \ e^{{i \over 2}k_{ls} \theta k_l}
\ <{\rm Tr}(\hat W_{C_2}[y_2]) \cdots {\rm Tr}(\hat W_{C_l}[y_l]_{0s}\hat W_{C_1}[y_1]
\hat W_{C_l}[y_l]_{s1}) \cdots {\rm Tr}(\hat W_{C_n}[y_n])>  \nonumber \\
&+& {i \over (2\pi)^4 {\rm det}\theta}\int_{C_1}dy_{1\mu}(s) 
\ e^{{i \over 2}k_{1s} \theta k_1} <{\rm Tr}(\hat W_{C_1}[y_1]_{0s}) 
{\rm Tr}(\hat W_{C_1}[y_1]_{s1}){\rm Tr}(\hat W_{C_2}[y_2]) \cdots {\rm Tr}(\hat W_{C_n}[y_n])>.
\nonumber \\
\label{threethree}
\eea
In this equation $y_l(s)-y_l(0)=\theta k_{ls}$ and each of the
contours $C_1, C_2, \cdots C_n$ may be open or closed.

\subsection{Closed Wilson loop}

Let us first consider a single closed Wilson loop. In this case
equation (\ref{threethree}) reduces to
\bea
{1 \over g^2} <{\rm Tr}(\hat W_C[y]\hat D_\nu\hat F_{\mu\nu}(\hat x))>
={i \over (2\pi)^4 {\rm det}\theta}\int_C \ dy_\mu(s) 
<{\rm Tr}(\hat W_C[y]_{0s}) {\rm Tr}(\hat W_C[y]_{s1})>
\label{threefour}
\eea
Here $C$ is a closed curve. The right hand side of (\ref{threefour})
has a disconnected piece. However, because of momentum conservation,
the disconnected piece contributes only when $y(s)=y(0)$. In fact,
both $<{\rm Tr}(\hat W_C[y]_{0s})>$ and $<{\rm Tr}(\hat W_C[y]_{s1})>$ are
proportional to $(2\pi)^4 \delta^{(4)}(k_s)$. One of these gives rise to
the total space-time volume $V$, while the other factor can be
rewritten as $(2\pi)^4 {\rm det}\theta \ \delta^{(4)}(y(s)-y(0))$. Thus,
we may rewrite (\ref{threefour}) as
\bea
&&{1 \over g^2} <{\rm Tr}(\hat W_C[y]\hat D_\nu\hat F_{\mu\nu}(\hat x))> \nonumber \\
&=& {i \over V} \int_C \ dy_\mu(s) \ \delta^{(4)}(y(s)-y(0))
<{\rm Tr}(\hat W_C[y]_{0s})> <{\rm Tr}(\hat W_C[y]_{s1})> \nonumber \\
&& + {i \over (2\pi)^4 {\rm det}\theta}\int_C \ dy_\mu(s)  
<{\rm Tr}(\hat W_C[y]_{0s}) {\rm Tr}(\hat W_C[y]_{s1})>_{\rm conn.}
\label{threefive}
\eea

The second term on the right hand side of (\ref{threefive}) contains
the connected part of the $2$-point function of open Wilson lines. At
finite $N$, it is easy to see that this term is down by a factor of
$1/N^2$ relative to the other terms in the equation. In the planar
limit, therefore, this term drops out and the planar equation looks
formally like the correponding equation in commutative gauge theory.
This is consistent with the perturbative result that, except for in an
overall phase, the dependence on the non-commutative parameter
$\theta$ drops out of planar diagrams. However, there are new
gauge-invariant observables in non-commutative gauge theory, the open
Wilson lines, and so there are new equations. As we shall see shortly,
these new equations have a non-trivial planar limit. One might then
say that it is these new equations that reflect the new physics of
non-commutative gauge theory.

At finite $N$ the second term on the right hand side of
(\ref{threefive}) contributes. One might wonder whether this term
reduces to its commutative counterpart in the limit of small
non-commutative parameter. An argument has been presented in \cite{AZD}
suggesting that this is the case. However, we find that, in fact, the
small $\theta$ limit of this term is not smooth, at least in
perturbation theory, as we will now show.

At the lowest order in perturbation theory, the second term on the
right hand side of (\ref{threefive}) evaluates to
\bea
{-ig^2N \over (2\pi)^4 {\rm det}\theta}\int_C dy_\mu(s) {1 \over k_s^2}
\bigg[(\int_0^s dy(\sigma) \ e^{-ik_s.y(\sigma)}).
(\int_s^1 dy(\sigma') \ e^{ik_s.y(\sigma')})\bigg].
\label{threesix}
\eea
For simplicity, let us specialize to a rectangular contour of sides
$L_1$ and $L_2$. We will also take $\theta$ to be of the form in
(\ref{twoonea}) with $\theta_a=\theta_b=\theta_0$. In this case the
diagrams that contribute to (\ref{threesix}) are shown in Fig. 2.

\begin{center}
\begin{picture}(450,125)(0,0)
\ArrowLine(50,25)(50,100)
\ArrowLine(50,100)(200,100)
\ArrowLine(200,100)(200,25)
\ArrowLine(100,25)(50,25)
\ArrowLine(197,25)(103,25)
\ArrowLine(250,25)(250,100)
\ArrowLine(250,100)(400,100)
\ArrowLine(400,100)(400,25)
\ArrowLine(300,25)(250,25)
\ArrowLine(397,25)(303,25)
\GlueArc(100,25)(30,0,180){3}{8}
\Gluon(325,100)(325,25){3}{8}
\Text(102,28)[b]{$s$}
\Text(47,60)[rb]{$L_2$}
\Text(150,22)[lt]{$L_1$}
\Text(197,22)[lt]{$\sigma=0$}
\Text(203,28)[lb]{$\sigma=1$}
\Text(302,28)[b]{$s$}
\Text(247,60)[rb]{$L_2$}
\Text(350,22)[lt]{$L_1$}
\Text(397,22)[lt]{$\sigma=0$}
\Text(403,28)[lb]{$\sigma=1$}
\end{picture} \\ 
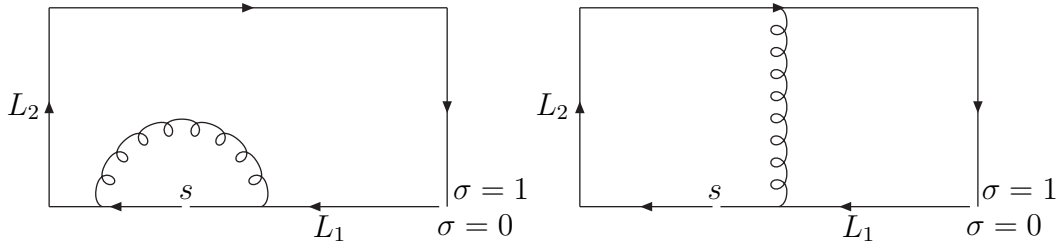
{\sl Fig. 2: Lowest order diagrams contributing to the non-planar term}
\end{center}

We can easily evaluate (\ref{threesix}) in this case. The result is
\bea
{g^2N \over (2\pi)^4 \theta_0^2} i(L_{1\mu}-L_{2\mu}) \bigg[1-\phi^{-2}
(1-e^{-i\phi})^2-f(\phi)+i\phi^{-1}e^{-i\phi}(f(\phi)-f^*(\phi))
\bigg]  
\label{threeseven}
\eea
where $\phi=L_1\theta^{-1}L_2$ is the magnetic flux passing through
the rectangular contour and 
\bea
f(\phi)=\int {ds \over s} \ (1-e^{-is\phi}).
\label{threeeight}
\eea 
In the limit of small non-commutativity, $\phi$ is large, and then
$f(\phi) \sim {\rm ln}\phi$. In this case the right hand side of
(\ref{threefive}) is divergent with the leading term going as $\sim
{\rm ln}\phi/\theta_0^2$. So we see that if we take the limit of small
non-commutativity first, keeping $N$ finite, we do not recover the
commutative result. It is easy to see in perturbation theory that the
origin of this problem lies in UV-IR mixing. It has been argued in
\cite{MRS} that this phenomenon renders loop diagrams finite in
non-commutative field theory. Now, the diagrams in Fig. 2 that
contribute to the right hand side of (\ref{threefive}) at order
$1/N^2$ in the lowest order in 't Hooft perturbation theory actually
come from non-planar one-loop diagrams on the left hand side of this
equation, as shown in Fig. 3.

\begin{center}
\begin{picture}(450,150)(0,0)
\ArrowLine(50,50)(50,125)
\ArrowLine(50,125)(200,125)
\ArrowLine(200,125)(200,50)
\ArrowLine(200,50)(50,50)
\ArrowLine(250,50)(250,125)
\ArrowLine(250,125)(400,125)
\ArrowLine(400,125)(400,50)
\ArrowLine(400,50)(250,50)
\GlueArc(100,50)(30,0,180){3}{8}
\GlueArc(150,100)(70,225,315){3}{12}
\Gluon(350,125)(350,50){3}{8}
\GlueArc(350,100)(70,225,315){3}{12}
\Text(100,53)[b]{$s$}
\Text(47,85)[rb]{$L_2$}
\Text(150,47)[lt]{$L_1$}
\Text(300,53)[b]{$s$}
\Text(247,85)[rb]{$L_2$}
\Text(350,47)[lt]{$L_1$}
\end{picture} \\
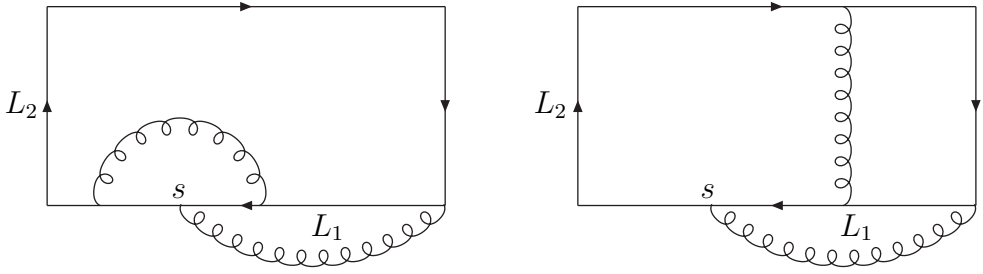
{\sl Fig. 3:  Non-planar one-loop diagrams giving rise to diagrams in Fig. 2.}
\end{center}

\ni In fact, the relevant amplitude is
\bea
g^4N \int dy_1\cdot\int dy_3 \int dy_2\cdot\int dy_4
\int {d^4p \over (2\pi)^4}\int {d^4q \over (2\pi)^4}
{e^{ip\cdot (y_1-y_3)}\over p^2}
{e^{iq\cdot (y_2-y_4)}\over q^2}e^{ip\theta q}
\label{threeeighta}
\eea
We can estimate the above momentum integral as follows
\bea
&&\int {d^4p\over (2\pi)^4}
\int {d^4q \over (2\pi)^4}{e^{ip\cdot (y_1-y_3)}\over p^2}
{e^{iq\cdot (y_2-y_4)}\over q^2}e^{ip\theta q} \nonumber \\
&=&
\left\{ \begin{array}{ll}
{1\over 4\pi^2}{1\over (y_1-y_3)^2}
{1\over 4\pi^2}{1\over (y_2-y_4)^2}
& \mbox{if $|y_1-y_3| |y_2-y_4| > \theta_0$} \\
{1\over 2(2\pi)^4}
{1\over \theta_0^2}{\rm ln}({\theta\over |y_1-y_3||y_2-y_4|})
& \mbox{if $|y_1-y_3| |y_2-y_4| < \theta_0$}
\end{array}
\right.
\eea
In commutative gauge theory these diagrams have short distance
singularities which are linearly divergent. In the non-commutative
theory they get regularized at the non-commutativity scale, as can be
seen from the above expression. The singularities of the commutative
theory reappear in the limit of small non-commutativity.  This is what
is reflected in the singular behaviour of the right hand side of
(\ref{threefive}) for small non-commutativity.

\subsection{Open Wilson lines}

The generic equation satisfied by the $n$-point function of Wilson
lines is (\ref{threethree}). The second term on the right hand side of
this equation has a disconnected part which is given by
\bea 
&&<{\rm Tr}(\hat W_{C_1}[y_1]_{0s})> <{\rm Tr}(\hat W_{C_1}[y_1]_{s1}){\rm Tr}(\hat
W_{C_2}[y_2]) \cdots {\rm Tr}(\hat W_{C_n}[y_n])> \nonumber \\
&+& <{\rm Tr}(\hat
W_{C_1}[y_1]_{s1})><{\rm Tr}(\hat W_{C_1}[y_1]_{0s}){\rm Tr}(\hat W_{C_2}[y_2])
\cdots {\rm Tr}(\hat W_{C_n}[y_n])>.
\label{threenine}
\eea
Because of momentum conservation, the first term contributes only for
$y(s)=y(0)$, while the second term contributes only for $y(s)=y(1)$.
In either term we get back the original $n$-point function. This is
just like for the closed Wilson loop discussed above. The connected
part of the second term on the right hand side can be easily seen to
be down by a factor of $1/N^2$ compared to the other terms in the
equation. In the planar limit, therefore, it drops out, leaving only
the ``joining'' term (the first term on the right hand side), apart
from the disconnected term mentioned above. We then have the result
that the planar Schwinger-Dyson equation for Wilson lines expresses
any $n$-point function entirely in terms of $(n-1)$-point functions.
By iterating this procedure $(n-1)$ times we may, in principle,
express any $n$-point function entirely in terms of {\it closed}
Wilson loops. 

The simplest example of the above phenomenon is provided by the
$2$-point function. In this case, the planar Schwinger-Dyson equation
reads
\bea
&&{1 \over g^2} <{\rm Tr}(\hat W_{C_1}[y_1]\hat D_\nu\hat F_{\mu\nu}(\hat x))
{\rm Tr}(\hat W_{C_2}[y_2])> \nonumber \\
&=& i\int_{C_2}dy_{2\mu}(s) \ e^{{i \over 2}k_{2s} \theta k_2}
\ <{\rm Tr}(\hat W_{C_2}[y_2]_{0s}\hat W_{C_1}[y_1]
\hat W_{C_2}[y_2]_{s1})>  \nonumber \\
&+& {i \over (2\pi)^4 {\rm det}\theta}\int_{C_1}dy_{1\mu}(s) 
\ e^{{i \over 2}k_{1s} \theta k_1} \bigg[<{\rm Tr}(\hat W_{C_1}[y_1]_{0s})> 
<{\rm Tr}(\hat W_{C_1}[y_1]_{s1}){\rm Tr}(\hat W_{C_2}[y_2])>
\nonumber \\
&& \hspace{20 ex}  +<{\rm Tr}(\hat
W_{C_1}[y_1]_{s1})><{\rm Tr}(\hat W_{C_1}[y_1]_{0s}){\rm Tr}(\hat W_{C_2}[y_2])> \bigg]. 
\label{threeten}
\eea 
We see that the right hand side involves closed Wilson loops, apart
from the the $2$-point function itself. The closed curves involved are
obtained by first traversing the curve $C_1$ given by $y_1(\sigma), \ 0
\leq \sigma \leq 1$ and then the curve given by
\bea
y(\sigma) &=& y_2(\sigma+s)-y_2(s)+y_1(1), \ \ 0 \leq \sigma \leq (1-s) 
\nonumber \\
&=& y_2(\sigma-1+s)-y_2(s)+y_1(0), \ \ (1-s) \leq \sigma \leq 1
\label{threeeleven}
\eea 
for different values of $s, \ 0 \leq s \leq 1$. Note that the closed
curves obtained in this way are continuous because of momentum
conservation.

Similarly, the $3$-point function involves two different $2$-point functions,
\bea
&&<{\rm Tr}(\hat W_{C_2}[y_2]_{0s}\hat W_{C_1}[y_1]
\hat W_{C_2}[y_2]_{s1}){\rm Tr}(\hat W_{C_3}[y_3])>, \nonumber \\ 
&&<{\rm Tr}(\hat W_{C_3}[y_3]_{0s}\hat W_{C_1}[y_1]
\hat W_{C_3}[y_3]_{s1}){\rm Tr}(\hat W_{C_2}[y_2])>,
\label{threetwelve}
\eea
depending on whether $C_1$ and $C_2$ or $C_1$ and $C_3$ combine into a
single curve. These $2$-point functions are themselves related to
closed loops, as discussed above. Thus, the $3$-point function can
eventually be related to closed Wilson loops, there being two distinct
sets of structures for the closed contours involved. These closed contours
can be obtained explicitly, as we have done above for the case of the
$2$-point function. For $n$-point function one eventually gets
$(n-1)!$ distinct structures for the closed loops.

The above discussion establishes a general link between the
correlators of Wilson lines and the expectation value of closed Wilson
loops. In a previous work \cite{DK2} we have presented a perturbative
proof for long straight Wilson lines to be bound together into closed
loops. The connection we have found here is more general in that the
Wilson lines are based on arbitrary contours and the momenta need not
be large. The closed loops we now find are also more general. We
believe that the planar Schwinger-Dyson equation indeed supports our
previous claim for the high energy behaviour of the Wilson lines. This
is because, firstly, it can be argued that the planar approximation is
always valid in the high energy limit (or large $\theta$ limit) since
the splitting term is supressed by $1/{\rm det}\theta$.  Secondly, we
need to repeatedly insert the equation of the motion operator into the
Wilson line correlator in order to eventually relate it to Wilson
loops. Such an operation picks up contact terms in the sense that it
makes two Wilson lines touch each other. In high energy limit, we
expect to rediscover such contact terms, although the real
singularities are expected to be regulated by the noncommutativity.
Finally, the set of relevant closed loops may collapse to that found
in \cite{DK2}, which is the set of extreme configurations of Wilson
loops obtained by simply joining the Wilson lines end-to-end.

\section{Loop equation for Wilson lines}

In this section we will first derive the loop equation for multipoint
correlators of Wilson lines. We will then consider the case of the
$2$-point function in detail and verify the planar loop equation in
this case upto second order in 't Hooft perturbation theory.

The loop equation is basically the Schwinger-Dyson equation
(\ref{threethree}) with the insertion of the equation of motion
operator replaced by a geometric variation of the contour. This is
done with the help of the identity
\bea
{\delta^2 \hat W_C[y] \over \delta y_\mu(\tau) \delta y_\mu(\tau')} &=&
e^{{i \over 2}k_{\tau'} \theta \tilde k_\tau} \ 
\hat W_C[y]_{0\tau}\bigg(i\del_\tau y_\nu(\tau) \hat F_{\mu\nu}(\hat x)\bigg)
\hat W_C[y]_{\tau\tau'}\bigg(i\del_{\tau'} y_\rho(\tau') 
\hat F_{\mu\rho}(\hat x)\bigg)\hat W_C[y]_{\tau'1} \nonumber \\
&& -\delta(\tau-\tau')e^{{i \over 2}k_\tau \theta k} \ 
\hat W_C[y]_{0\tau}\bigg(i\del_\tau y_\mu(\tau) 
\hat D_\nu \hat F_{\mu\nu}(\hat x)\bigg)\hat W_C[y]_{\tau 1}
\label{fourone}
\eea
where, as before, $\theta k_\tau=y(\tau)-y(0)$ and
$\tilde k_{\tau}=k-k_\tau$. Note that this identity is valid at interior
points of the Wilson line. At the boundaries of the Wilson line one
has to be more careful. However, if we vary {\it both} the ends
keeping $k$ fixed, and assume that the tangents to the contour at the
ends are identical, \footnote{Under these conditions the variation of
contour at the end points is effectively like at an interior point.} 
then a very similar identity is valid at the ends also.

We need to separate out the equation of motion piece on the right hand
side of (\ref{fourone}). This may formally be done by defining the
``loop laplacian'' \cite{POLY}
\bea
{\del^2 \over \del y^2(\tau)}\equiv {\rm lim}_{\epsilon \rightarrow 0} 
\int^\epsilon_{-\epsilon}dt \ {\delta^2 \over \delta y_\mu(\tau+t/2) 
\delta y_\mu(\tau-t/2)}.
\label{fourtwo}
\eea 
In principle, if the quantum theory is regularized then the first term
in (\ref{fourone}) does not have any singularities as $\tau
\rightarrow \tau'$ and so the loop laplacian picks up only the
delta-function term on the right hand side of this equation.
\footnote{In practice the separation of the two terms on the right
hand side of (\ref{fourone}) is a nontrivial issue. For a discussion
in the case of commutative gauge theory, see, for example, 
\cite{POLY,PR,DGO}.} We then get
\bea
-{\del^2 \hat W_C[y] \over \del y^2(\tau)}=e^{{i \over 2}k_\tau \theta k}
\ \hat W_C[y]_{0\tau}\bigg(i\del_\tau y_\mu(\tau) 
\hat D_\nu \hat F_{\mu\nu}(\hat x)\bigg)\hat W_C[y]_{\tau 1}.
\label{fourthree}
\eea
Using this in (\ref{threethree}) we get the loop equation for Wilson
line correlators
\bea
&& -{1 \over g^2}{\del^2 \over \del y_1^2(\tau)}<W_{C_1}[y_1]W_{C_2}[y_2]
\cdots W_{C_n}[y_n]> \nonumber \\
&=& \sum^n_{l=2} \int_{C_l} \ ds \ (i\del_\tau y_1(\tau).i\del_s y_l(s)) 
\ e^{-ik_1.y_1(\tau)-ik_l.y_l(s)} \nonumber \\
&& \hspace{5 ex} \times <W_{C_2}[y_2] \cdots  
{\rm Tr}(\hat W_{C_{1\tau}}[y_{1\tau}] 
\hat W_{C_{ls}}[y_{ls}]) \cdots W_{C_n}[y_n]> \nonumber \\
&& + {1 \over (2\pi)^4 {\rm det}\theta} \int_{C_{1\tau}} \ ds \ 
(i\del_\tau y_1(\tau).i\del_s y_1(s)) \ e^{-ik_1.y_1(\tau) 
+{i \over 2}k_{1s} \theta k_1} \nonumber \\
&& \hspace{15 ex} \times <{\rm Tr}(\hat W_{C_{1\tau}}[y_{1\tau}]_{0s})
{\rm Tr}(\hat W_{C_{1\tau}}[y_{1\tau}]_{s1}) W_{C_2}[y_2] \cdots W_{C_n}[y_n]>,
\label{fourfour}
\eea
where the contours $C_{1\tau}$ and $C_{1s}$ are as defined in
(\ref{twothirteen}). Also, the operator $\hat W_C[y]$ has been defined
in (\ref{twoeight}) and $W_C[y]$ is the gauge-invariant observable
defined in (\ref{oneone}).

\subsection{Two-point function}

We will now discuss the case of the $2$-point function in some detail.
For the $2$-point function, the loop equation reduces to
\bea
&& -{1 \over g^2}{\del^2 \over \del y_1^2(\tau)}<W_{C_1}[y_1]W_{C_2}[y_2]>
\nonumber \\
&=& \int_{C_2} \ ds \ (i\del_\tau y_1(\tau).i\del_s y_2(s)) 
\ e^{-ik_1.y_1(\tau)-ik_2.y_2(s)}
<{\rm Tr}(\hat W_{C_{1\tau}}[y_{1\tau}] \hat W_{C_{2s}}[y_{2s}])> 
\nonumber \\
&& + {1 \over (2\pi)^4 {\rm det}\theta} \int_{C_{1\tau}} \ ds \ 
(i\del_\tau y_1(\tau).i\del_s y_{1\tau}(s)) \ e^{-ik_1.y_1(\tau) 
+{i \over 2}k_{1s} \theta k_1} \nonumber \\
&& \hspace{20 ex} <{\rm Tr}(\hat W_{C_{1\tau}}[y_{1\tau}]_{0s})
{\rm Tr}(\hat W_{C_{1\tau}}[y_{1\tau}]_{s1}) W_{C_2}[y_2]>.
\label{fourfive}
\eea

We are interested in the planar limit of this equation. In this limit
only the disconnected part of the $3$-point function, appearing
on the right hand side of (\ref{fourfive}), survives. This
disconnected part is $$<{\rm Tr}(\hat W_{C_{1\tau}}[y_{1\tau}]_{0s})>
<{\rm Tr}(\hat W_{C_{1\tau}}[y_{1\tau}]_{s1}) W_{C_2}[y_2]>+<{\rm Tr}(\hat
W_{C_{1\tau}}[y_{1\tau}]_{s1})>< {\rm Tr}(\hat
W_{C_{1\tau}}[y_{1\tau}]_{0s}) W_{C_2}[y_2]>.$$
Because of momentum conservation, the first term survives only when 
$y_{1\tau}(s)=y_{1\tau}(0)$, while the second term survives only 
when $y_{1\tau}(s)=y_{1\tau}(1)$. Assuming that the contour $C_{1\tau}$
has no self-intersection, the first condition is satisfied only for 
$s=0$, while the second condition is satisfied only for $s=1$. Thus 
the disconnected part of the $3$-point function takes the form   
$$(<{\rm Tr}(\hat W_{C_{1\tau}}[y_{1\tau}]_{0s})>+<{\rm Tr}(\hat
W_{C_{1\tau}}[y_{1\tau}]_{s1})>)<{\rm Tr}(\hat W_{C_{1\tau}}[y_{1\tau}]) 
W_{C_2}[y_2]>,$$ which, using (\ref{twotwelve}), is equivalent to
$$ e^{ik_1.y_1(\tau)} \
(<{\rm Tr}(\hat W_{C_{1\tau}}[y_{1\tau}]_{0s})>+<{\rm Tr}(\hat
W_{C_{1\tau}}[y_{1\tau}]_{s1})>)<W_{C_1}[y_1] W_{C_2}[y_2]>.$$ 
Using this in (\ref{fourfive}), together with the fact that $\del_\tau
y_\mu(\tau)$ gives tangents at the two ends of the contour $C_{1\tau}$,
which are equal by construction, we get 
\bea
&& -{1 \over g^2}{\del^2 \over \del y_1^2(\tau)}<W_{C_1}[y_1]W_{C_2}[y_2]>
\nonumber \\
&=& -{(\del_\tau y_1(\tau))^2 \over (2\pi)^4 {\rm det}\theta} 
\int_{C_{1\tau}} \ ds \ 
(<{\rm Tr}(\hat W_{C_{1\tau}}[y_{1\tau}]_{0s})>+
<{\rm Tr}(\hat W_{C_{1\tau}}[y_{1\tau}]_{s1})>)<W_{C_1}[y_1] W_{C_2}[y_2]>
\nonumber \\
&& +\int_{C_2} \ ds \ (i\del_\tau y_1(\tau).i\del_s y_2(s)) 
\ e^{-ik_1.y_1(\tau)-ik_2.y_2(s)} 
<{\rm Tr}(\hat W_{C_{1\tau}}[y_{1\tau}] \hat W_{C_{2s}}[y_{2s}])>. 
\label{foursix}
\eea
This is the final form of the planar loop equation for the $2$-point
function.  Notice that the first term on the right hand side of this
equation is proportional to $(\del_\tau y_1(\tau))^2$ and also it
involves the original $2$-point function. Taken together with the left
hand side, the two terms have the form of string hamiltonian acting on
the $2$-point function. However, it is not clear that this term is
really physically meaningful. In fact, a corresponding term in the
loop equation for the commuting gauge theory is often ignored in the
regularized theory. In the present case also, an evaluation of the
coefficient of $(\del_\tau y_1(\tau))^2$ cannot be done unambiguously.
This is because such a calculation involves computation of amplitudes
for splitting off of tiny bits at the two ends of the Wilson line
defined on the contour $C_{1\tau}$. The computation of this is
delicate and needs a regulator. So the physical significance of this
term remains unclear.

We should mention here that equations (\ref{fourfour}) and
(\ref{foursix}) are new type of loop equations since there is no
analogue of these in commutative gauge theory. Also, it is clear that
the planar equation (\ref{foursix}) relates the $2$-point function of
open Wilson lines to the expectation value of a closed Wilson loop.
The latter may be obtained by solving the planar loop equation for a
closed Wilson loop. Thus one needs equations for both types of Wilson
observables to form a closed system of equations.

\subsection{Perturbative verification}

Perhaps the most interesting aspect of the loop equation
(\ref{foursix}) is its stringy interpretation. Investigating this
aspect of the loop equation is bound to be inherently
non-perturbative. In fact, recently such an exercise has been
successfully carried out in \cite{PR,DGO} for the loop equation in
commutative gauge theory, using the AdS/CFT correspondence. A similar
exercise for the present non-commutative case seems to require a
better understanding of the connection between non-commutative gauge
theory and its conjectured gravity dual \cite{MR,HI}, and is beyond the
scope of the present work. Here we will restrict ourselves to a
perturbative verification of (\ref{foursix}).  We will do the
computations upto the second order in the 't Hooft coupling.

Separating out the momentum conserving delta-function, we may
parametrize the $2$-point function as
\bea
<W_{C_1}[y_1] W_{C_2}[y_2]>=(2\pi)^4 \delta^{(4)}(k_1+k_2) \ 
G_{C_1C_2}[y_1, y_2],
\label{fourseven}
\eea
where, in perturbation theory,
\bea
G_{C_1C_2}[y_1, y_2]=\lambda G^{(1)}_{C_1C_2}[y_1, y_2]+\lambda^2
G^{(2)}_{C_1C_2}[y_1, y_2]+ \cdots.
\label{foureight}
\eea
Here $\lambda=g^2N$ is the 't Hooft coupling constant. As indicated
in (\ref{foureight}), in perturbation theory the function
$G_{C_1C_2}[y_1, y_2]$ starts at first order in the 't Hooft coupling
and is order one in $N$. The left hand side of (\ref{foursix}) is,
therefore, of order $N$, the same as the right hand side.

The lowest order diagram contributing to the $2$-point function is
shown in Fig. 4.

\begin{center}
\begin{picture}(400,100)(0,0)
\ArrowArc(200,50)(50,120,240)
\ArrowArc(200,50)(50,300,60)
\Gluon(157,75)(243,25){3}{16}
\Text(147,50)[r]{$C_1$}
\Text(255,50)[l]{$C_2$}
\end{picture} \\
{\sl Fig. 4: Lowest order diagram contributing to the $2$-point function.}
\end{center}

\ni A simple calculation gives the result 
\bea
G^{(1)}_{C_1C_2}[y_1, y_2]={1 \over k_1^2}V_{C_1}^{(k_1)}[y_1].
V_{C_2}^{(k_2)}[y_2].
\label{fournine}
\eea
where $V_C^{(k)}[y]$ has been defined in (\ref{twofourteen}). Operating 
the loop laplacian on (\ref{fournine}), we get the lowest order expression 
for the left hand side of (\ref{foursix}) \footnote{Here and in the
following we have omitted the momentum conserving delta-function
factor $(2\pi)^4 \delta^{(4)}(k_1+k_2)$ which is present on both
sides of (\ref{foursix}).}
\bea
\bigg(i\del_\tau y_{1\mu}(\tau) e^{-ik_1.y_1(\tau)} \bigg)
\ V_{C_2\mu}^{(k_2)}[y_2].
\label{fourten}
\eea
In arriving at this expression we have used that $k.V_C^{(k)}[y]=0$,
which is true because of the identity $k.y(1)=k.y(0)$ which follows
from the definition of $k$, (\ref{onethree}).

On the right hand side of (\ref{foursix}), at the lowest order in
$\lambda$, the first term does not contribute. The relevant
contribution comes from the second term by setting the gauge field
to zero in each of the two Wilson line operators involved in making
the closed Wilson loop. Omitting the momentum conserving
delta-function, we get precisely the expression in (\ref{fourten}). 

At the next order in $\lambda$, there are several different types of
diagrams that contribute to the $2$-point function. Fig. 5 shows a
representative example from each type. 

\begin{center}
\begin{picture}(450,125)(0,0)
\ArrowArc(90,75)(50,120,240)
\ArrowArc(60,75)(50,300,60)
\Text(77,12)[b]{(a)}
\Text(40,75)[r]{$C_1$}
\Text(112,75)[l]{$C_2$}
\GlueArc(47,100)(15,53,247){3}{5}
\Gluon(50,45)(100,45){3}{5}
\ArrowArc(190,75)(50,120,240)
\ArrowArc(160,75)(50,300,60)
\Text(177,12)[b]{(b)}
\Gluon(150,45)(200,45){3}{5}
\Gluon(150,105)(200,105){3}{5}
\ArrowArc(290,75)(50,120,240)
\ArrowArc(260,75)(50,300,60)
\Text(277,12)[b]{(c)}
\Gluon(250,105)(275,75){3}{4}
\Gluon(250,45)(275,75){3}{4}
\Gluon(275,75)(310,80){3}{3}
\Vertex(275,75){2}
\ArrowArc(390,75)(50,120,240)
\ArrowArc(360,75)(50,300,60)
\Text(377,12)[b]{(d)}
\Gluon(390,80)(410,80){3}{2}
\Gluon(340,80)(360,80){3}{2}
\GlueArc(375,80)(15,0,180){3}{4}
\GlueArc(375,80)(15,180,360){3}{4}
\Vertex(390,80){2}
\Vertex(360,80){2}
\end{picture} \\ 
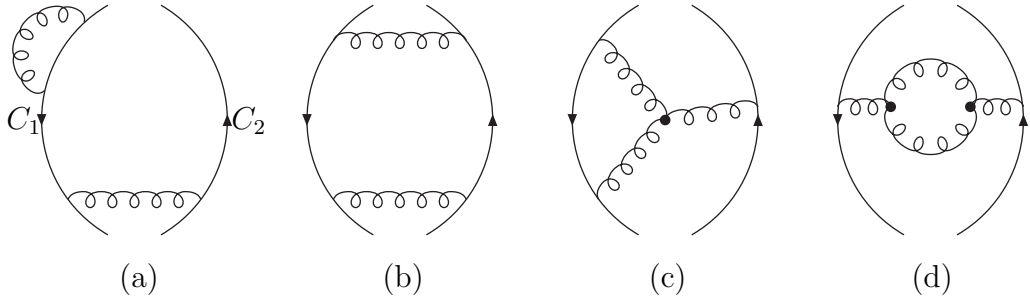
{\sl Fig. 5: Examples of diagrams contributing to the $2$-point function at 
second order in perturbation theory.}
\end{center}

\ni We have done a calulation of the quantity $G^{(2)}_{C_1C_2}[y_1,
y_2]$, which gives the second order contribution to the $2$-point
function. Some details of this calculation and the result are given in
the Appendix.  

At the second order in $\lambda$, both terms on the right hand side of
(\ref{foursix}) contribute. The contribution of the first term comes
from the lowest order calculation of the $2$-point function, while
that of the second term comes from a one gauge boson exchange. In the
Appendix we have discussed in detail how each of these contributions
arises as a result of operating the loop laplacian on
$G^{(2)}_{C_1C_2}[y_1, y_2]$. Here we only mention that some of the
terms from different sets of diagrams that appear in the calculation
of the left hand side of (\ref{foursix}) have a structure that does
not occur on the right hand side. However, there are non-trivial
cancellations between the contributions of different sets of diagrams.
We have checked that many such terms disappear from the overall result
for the left hand side as well, but we have not attempted a complete
verification of this.

It would be interesting to extend the present perturbative analysis to
all orders in $\lambda$. To verify the new loop equation
(\ref{fourfour}) non-perturbatively, one needs to understand what the
multipoint correlators of Wilson lines based on arbitrary contours map
on to in the string/supergravity dual. A better understanding of the
non-commutative gauge theory/string theory duality than we have at
present appears to be necessary for this.

\section{Discussion}

In this paper we have investigated Schwinger-Dyson and loop equations in
non-commutative gauge theory. A major difference from the commutative
case is the existence of gauge-invariant Wilson line observables based
on open contours, in addition to those on closed contours. The
Schwinger-Dyson and loop equations in non-commutative gauge theories,
therefore, involve both types of gauge-invariant Wilson observables.
In the planar limit, the equations for a closed Wilson loop simplify
and, like in their commutative counterparts, involve only closed
loops.  There are, however, {\it new} equations, those for open Wilson
lines. These involve closed Wilson loops as well, so both types of
Wilson observables are needed for a closed set of equations in the
planar limit. In fact, as we have seen, these latter equations
determine correlators of open Wilson lines entirely in terms of closed
Wilson loops.

Recently in several works it has been argued \cite{GHI,HL,DT,OO,LM,O} 
that local operators in non-commutative gauge theory with straight Wilson
lines attached to them are dual to bulk supergravity modes. In this
context it is relevant to ask what bulk observables are dual to Wilson
lines based on arbitrary open contours. This question is also
important for a non-perturbative study of the new loop equations
derived here. Our proposal is to identify a Wilson line based on an
arbitrary open contour with the operator dual to bulk {\it closed}
string. This proposal is based on the following reasoning.

The momentum variable appearing in a Wilson line satisfies the
condition (\ref{onethree}). This is a constraint on the contour
enforced by gauge invariance. The contour is otherwise arbitrary. This
condition may be regarded as a boundary condition on the curves
involved. A generic curve with this boundary condition may be
parametrized as $y(\sigma)=\sigma(\theta k)+y'(\sigma)$, where $0 \leq
\sigma \leq 1$ and $y'(\sigma)$ satisfies {\it periodic} boundary
conditions. Thus the freedom contained in a generic Wilson line is
exactly the one needed to describe a closed string!

Actually, we can take this line of reasoning further. Let us confine
our attention to smooth curves, with the additional condition that the
tangents to the curve at the two ends are equal. In this case we may
parametrize the curves as
\bea
y(\sigma) &=& y_0(\sigma)+\delta y(\sigma), \nonumber \\
y_0(\sigma) &=& y_0(0)+\sigma(\theta k), \nonumber \\
\delta y(\sigma) &=& \sum_{n=1}^\infty (\alpha_n \ e^{-2\pi i n \sigma}+
\tilde \alpha_n \ e^{2\pi i n \sigma}).
\label{fiveone}
\eea
As the above parametrization suggests, what we are going to do is to
assume that deviations of the given curve from a straight line are
small and expand the Wilson line, based on the given curve, around the
corresponding straight Wilson line. This gives
\bea
W_C[y] &=& W_{C_0}[y_0]+\int_0^1 d\sigma \ \delta y_\mu(\sigma) 
\bigg({\delta  W_C[y] \over \delta y_\mu(\sigma)} \bigg)_{y=y_0} \nonumber \\
&& +{1 \over 2!} \int_0^1 d\sigma \int_0^1 d\sigma' \ \delta y_\mu(\sigma) 
\ \delta y_\nu(\sigma') \bigg({\delta^2  W_C[y] \over \delta y_\mu(\sigma)
\delta y_\nu(\sigma')} \bigg)_{y=y_0}+\cdots
\label{fivetwo}
\eea
Here $C_0$ refers to the straight line contour. 

The first term in the above equation is known to be the
non-commutative gauge theory operator dual to the bulk closed string
tachyon. The second term vanishes, since $\bigg({\delta W_C[y] \over
  \delta y_\mu(\sigma)} \bigg)_{y=y_0}$ is independent of $\sigma$,
which can be easily verified using the cyclic symmetry of a straight
Wilson line, and since $\delta y(\sigma)$ has no zero mode. The first
non-trivial contribution comes from the third term. Using a
generalization of the identity in (\ref{fourone}), we may
rewrite this term as
\bea
&& \int_0^1 d\sigma \int_0^1 d\sigma' \ \delta y_\mu(\sigma) 
\ \delta y_\nu(\sigma') \ \bigg [ {\rm Tr} 
\bigg(\hat {\cal U}_{C_0}(0, \sigma) \ (il_\lambda 
\hat F_{\mu\lambda}(\hat x+y_0(\sigma))) \ 
\hat {\cal U}_{C_0}(\sigma, \sigma') \nonumber \\
&& \hspace{10 ex} \times (il_\rho \hat F_{\nu\rho}(\hat x+y_0(\sigma'))) \  
\hat {\cal U}_{C_0}(\sigma', 1) \ e^{ik.\hat x} \bigg) \theta(\sigma'-\sigma)+
(\sigma' \leftrightarrow \sigma, \mu \leftrightarrow \nu )\bigg ] \nonumber \\
&& + \int_0^1 d\sigma  \ \delta y_\mu(\sigma) \ \delta y_\nu(\sigma) \ 
{\rm Tr} \bigg(\hat {\cal U}_{C_0}(0, \sigma) \ (il_\sigma 
\hat D_\nu \hat F_{\mu\sigma}
(\hat x+y_0(\sigma))) \ \hat {\cal U}_{C_0}(\sigma, 1) \ e^{ik.\hat x}
\bigg) \nonumber \\
&& + \int_0^1 d\sigma \ \delta y_\mu(\sigma) \ i\del_\sigma
\delta y_\nu(\sigma) \ {\rm Tr} \bigg(\hat {\cal U}_{C_0}(0, \sigma) \ 
\hat F_{\mu\nu}(\hat x+y_0(\sigma)) \ 
\hat {\cal U}_{C_0}(\sigma, 1) \ e^{ik.\hat x} \bigg), \nonumber \\
\label{fivethree}
\eea 
where $\hat {\cal U}_{C_0}(\sigma_1, \sigma_2)$ is the path-ordered
phase factor, running along the straight line contour $C_0$, from the
point $\sigma_1$ to $\sigma_2$. Note that in this notation $W_C[y]=
{\rm Tr}(\hat {\cal U}_{C_0}(0, 1) \ e^{ik.\hat x})$.  Substituiting
for $\delta y(\sigma)$ from (\ref{fiveone}) in this expression and
extracting the part of the $n=1$ term symmetric in the indices $\mu,
\nu$, we get precisely the operator that has been identified in
\cite{OO} as being dual to the bulk graviton (in the bosonic string),
polarized along the brane directions, modulo factors that connect the
open string metric with the closed string metric and terms involving
the scalar fields. Note that the last term in (\ref{fivethree}) is
purely antisymmetric in the indices $\mu, \nu$ and hence contributes
only to the operator dual to the bulk antisymmetric tensor field.

It seems quite likely that the above procedure gives us {\it all} the
gauge theory operators dual to bulk string modes. It is, therefore,
tempting to identify the Wilson line based on generic curves of the
type described by (\ref{fiveone}) as dual to the bulk {\it closed}
string. An expansion of the Wilson line around the corresponding
straight line contour would then be like the expansion of the {\it
  closed string field} in terms of the various string modes carrying a
definite momentum. If this is true, then multipoint correlators of
Wilson lines should be identified with closed string scattering
amplitudes. In particular, the $2$-point function would then have the
interpretation of closed string propagator and (\ref{foursix}) would
be the equation of motion satisfied by the propagator. Such a
non-perturbative interpretation of (\ref{foursix}), or more generally
(\ref{fourfour}), should further enhance our understanding of gauge
theory/string theory duality. The above discussion applies to the
bosonic string. It would be interesting to extend these ideas to the
case of the superstring.

\newpage

\appendix
\section{Appendix}

In this appendix we will give some details of the calculation of
$G^{(2)}_{C_1C_2}[y_1, y_2]$. We will also describe how the action of
the loop laplacian on it reproduces the right hand side of the loop
equation (\ref{foursix}).

The diagrams that contribute to $G^{(2)}_{C_1C_2}[y_1, y_2]$ can be
collected into four different types of groups. A representative from
each of these has been shown in Fig. 5. There are six self-energy type
of diagrams, Fig. 5(a). Their total contribution to
$G^{(2)}_{C_1C_2}[y_1, y_2]$ is
\bea
&& {1 \over k_1^2} \int_{\sigma_1}\int_{\sigma_2>{\sigma_2}'>{\sigma_2''}} 
\bigg[ {(\dot y_1(\sigma_1).\dot y_2({\sigma_2}''))
(\dot y_2(\sigma_2).\dot y_2({\sigma_2}')) \over 
4 \pi^2 |y_2(\sigma_2)-y_2({\sigma_2}')|^2} \ 
e^{-ik_1.(y_1(\sigma_1)-y_2({\sigma_2}''))} \nonumber \\
&& \hspace{20 ex}  +{(\dot y_1(\sigma_1).\dot y_2({\sigma_2}'))
(\dot y_2(\sigma_2).\dot y_2({\sigma_2}'')) \over 
4 \pi^2 |y_2(\sigma_2)-y_2({\sigma_2}'')+\theta k_1|^2} \ 
e^{-ik_1.(y_1(\sigma_1)-y_2({\sigma_2}'))} \nonumber \\
&& \hspace{20 ex} +{(\dot y_1(\sigma_1).\dot y_2(\sigma_2))
(\dot y_2({\sigma_2}').\dot y_2({\sigma_2}'')) \over 
4 \pi^2 |y_2({\sigma_2}')-y_2({\sigma_2}'')|^2} \  
e^{-ik_1.(y_1(\sigma_1)-y_2(\sigma_2))} \bigg] \nonumber \\
&& +1 \leftrightarrow 2
\label{aone}
\eea
where a dot on $y$ stands for a derivative with respect to the
argument and the last contribution is obtained by the $1
\leftrightarrow 2$ interchange of the subscripts on $k$, $y$ and
$\sigma$.

There are two diagrams in the second set represented by Fig. 5(b). 
Their total contribution  to $G^{(2)}_{C_1C_2}[y_1, y_2]$ is
\bea
&& \int d^4z \ e^{ik_1.z} \int_{\sigma_1>{\sigma_1}'}
\int_{\sigma_2>{\sigma_2}'}
\bigg[ {(\dot y_1(\sigma_1).\dot y_2(\sigma_2))(\dot y_1({\sigma_1}').
\dot y_2({\sigma_2}')) \over 4 \pi^2 |z+y_1(\sigma_1)-y_2(\sigma_2)|^2 \
4 \pi^2 |z+y_1({\sigma_1}')-y_2({\sigma_2}')+\theta k_1|^2} \nonumber \\
&& \hspace{10 ex} +{(\dot y_1({\sigma_1}').\dot y_2(\sigma_2))
(\dot y_1(\sigma_1).\dot y_2({\sigma_2}')) 
\over 4 \pi^2 |z+y_1(\sigma_1)-y_2({\sigma_2}')|^2 \
4 \pi^2 |z+y_1({\sigma_1}')-y_2(\sigma_2)|^2} \bigg]. 
\label{atwo}
\eea

In the third set, represented by Fig. 5(c), also there are two
diagrams. Their total contribution to $G^{(2)}_{C_1C_2}[y_1, y_2]$ is
\bea
&& -{1 \over k_1^2} \int d^4z \int_{\sigma_1}\int_{\sigma_2>{\sigma_2}'} \ 
{e^{ik_1.(z+y_2(\sigma_2)-y_1(\sigma_1))} \over 4 \pi^2 z^2} 
\bigg[ -ik_{1\nu}\dot y_{1\mu}(\sigma_1) \bigg( \dot y_{2\mu}({\sigma_2}')
\dot y_{2\nu}(\sigma_2)-2\dot y_{2\mu}(\sigma_2)
\dot y_{2\nu}({\sigma_2}') \bigg) \nonumber \\
&& \hspace{15 ex} +\dot y_{1\mu}(\sigma_1)\bigg(\dot y_{2\mu}({\sigma_2}') 
\dot y_{2\nu}(\sigma_2)+\dot y_{2\mu}(\sigma_2)\dot y_{2\nu}({\sigma_2}')
-2\dot y_2(\sigma_2).\dot y_2({\sigma_2}')\delta_{\mu\nu} \bigg) 
{\del_z}_\nu \bigg] \nonumber \\
&& \hspace{20 ex} \times 
\bigg({1 \over 4 \pi^2 |z+y_2(\sigma_2)-y_2({\sigma_2}')+
\theta k_1|^2}-{1 \over 4 \pi^2 |z+y_2(\sigma_2)-y_2({\sigma_2}')|^2}
\bigg) \nonumber \\ && +1 \leftrightarrow 2
\label{athree}
\eea

Finally, we have the gauge boson self-energy diagrams like Fig. 5(d),
including those with ghosts. Their total contribution to
$G^{(2)}_{C_1C_2}[y_1, y_2]$ is
\bea
&& -{1 \over {(k_1^2)}^2} \int d^4z \ \int_{\sigma_1}\int_{\sigma_2} \
e^{ik_1.(z-y_1(\sigma_1)+y_2(\sigma_2))} 
{\dot y_{1\mu}(\sigma_1)\dot y_{2\nu}(\sigma_2) 
\over 4 \pi^2 z^2} \nonumber \\
&& \hspace{5 ex} \times \bigg( \delta_{\mu\nu}(-{\del_z}^2
+2i\delta_{\mu\nu}k_1.\del_z-5k_1^2)+8{\del_z}_\mu {\del_z}_\nu \bigg) 
\bigg( {1 \over 4 \pi^2 |z+\theta k_1|^2} -{1 \over 4 \pi^2 z^2} \bigg). 
\label{afour}
\eea

Let us now evaluate the action of the loop laplacian, $-\del^2/ \del
y_1^2(\tau)$, on the expression for the second order contribution to
the $2$-point function given above. In the first term in (\ref{aone}),
the only dependence on $y_1$ is in the form of
$V_{C_1\mu}^{(k_1)}[y_1]$, which has been defined in
(\ref{twofourteen}). Applying the loop laplacian on it gives the
result
\bea
-{\del^2 \over \del y_1^2(\tau)}V_{C_1\mu}^{(k_1)}[y_1]=
(k_1^2\delta_{\mu\nu}-k_{1\mu}k_{1\nu}) \dot y_{1\nu}(\tau) \ 
e^{-ik_1.y_1(\tau)}.
\label{afive}
\eea
The $k_1^2$ in the first term above cancels the factor of $1/k_1^2$ in
front of the first term in (\ref{aone}). The rest of this factor can
be seen to precisely reproduce that contribution of the second term on
the right hand side of the loop equation (\ref{foursix}) in which a
self-energy insertion is present on the contour $C_{2s}$. The three
terms correspond to the three possibilities that the marked point $s$
on the contour $C_{2s}$ is entirely above, entirely below or
in-between the points where the self-energy insertion takes place. The
second term in (\ref{afive}) gives rise to the following contribution
from the first term in (\ref{aone}):
\bea
&& -{k_1.\dot y_1(\tau) \over k_1^2}e^{-ik_1.y_1(\tau)}
\int_{\sigma_2>{\sigma_2}'>{\sigma_2''}} 
\bigg[ {(k_1.\dot y_2({\sigma_2}''))
(\dot y_2(\sigma_2).\dot y_2({\sigma_2}')) \over 
4 \pi^2 |y_2(\sigma_2)-y_2({\sigma_2}')|^2} \ 
e^{ik_1.y_2({\sigma_2}'')} \nonumber \\
&& +{(k_1.\dot y_2({\sigma_2}'))
(\dot y_2(\sigma_2).\dot y_2({\sigma_2}'')) \over 
4 \pi^2 |y_2(\sigma_2)-y_2({\sigma_2}'')+\theta k_1|^2} \ 
e^{ik_1.y_2({\sigma_2}')}+{(k_1.\dot y_2(\sigma_2))
(\dot y_2({\sigma_2}').\dot y_2({\sigma_2}'')) \over 
4 \pi^2 |y_2({\sigma_2}')-y_2({\sigma_2}'')|^2} \  
e^{ik_1.y_2(\sigma_2)} \bigg]. \nonumber \\
\label{asix}
\eea
This can be simplified to 
\bea
&& {ik_1.\dot y_1(\tau) \over k_1^2}e^{-ik_1.y_1(\tau)}
\int_{\sigma_2>{\sigma_2}'}(\dot y_2(\sigma_2).\dot y_2({\sigma_2}')) \ 
(e^{ik_1.y_2(\sigma_2)}-e^{ik_1.y_2({\sigma_2}')}) \nonumber \\
&& \hspace{10 ex} \times \bigg( 
{1 \over 4 \pi^2 |y_2(\sigma_2)-y_2({\sigma_2}')+\theta k_1|^2}
-{1 \over 4 \pi^2 |y_2(\sigma_2)-y_2({\sigma_2}')|^2} \bigg)
\label{aseven}
\eea
Now, it is easy to see that on the right hand side of (\ref{foursix}) 
there are no terms at this order having the above structure.
Therefore, for consistency of the loop equation, (\ref{aseven}) must
get cancelled by another term in the $2$-point function. In fact, this
does happen and the required term comes from the first term of
(\ref{athree}). In this term also $y_1$ is in the form of
$V_{C_1\mu}^{(k_1)}[y_1]$, so applying the loop laplacian results in
two terms because of (\ref{afive}). Let us look at the second term. It is
\bea
&& {k_1.\dot y_1(\tau) \over k_1^2}e^{-ik_1.y_1(\tau)}
\int d^4z \int_{\sigma_2>{\sigma_2}'} \ 
{e^{ik_1.(z+y_2(\sigma_2))} \over 4 \pi^2 z^2} 
\bigg[ -ik_1.\dot y_2({\sigma_2}')k_1.\dot y_2(\sigma_2) \nonumber \\
&& \hspace{10 ex} +k_{1\mu}\bigg(\dot y_{2\mu}({\sigma_2}') 
\dot y_{2\nu}(\sigma_2)+\dot y_{2\mu}(\sigma_2)\dot y_{2\nu}({\sigma_2}')
-2\dot y_2(\sigma_2).\dot y_2({\sigma_2}')\delta_{\mu\nu} \bigg) 
{\del_z}_\nu \bigg] \nonumber \\
&& \hspace{10 ex} \times 
\bigg({1 \over 4 \pi^2 |z+y_2(\sigma_2)-y_2({\sigma_2}')+
\theta k_1|^2}-{1 \over 4 \pi^2 |z+y_2(\sigma_2)-y_2({\sigma_2}')|^2} \bigg)
\label{aeight}
\eea
In the last term in the square brackets above, let us rewrite
$2k_1.\del_z$ as $-i\{(k_1+i\del_z)^2+\del_z^2-k_1^2\}$ and then use
$(k_1+i\del_z)^2(e^{ik_1.z}/z^2)=e^{ik_1.z}\del_z^2(1/z^2)$ and
$\del_z^2(1/z^2)=-4\pi^2\delta^{(4)}(z)$. As a result of this
simplification, one of the terms we get is precisely (\ref{aseven}),
but with opposite sign. So this unwanted term cancels in a rather
nontrivial way, since the cancellation involves terms which come from
two entirely different diagrams.

Going back to (\ref{aone}), let us now look at the second term, which
is obtained from the first by $1 \leftrightarrow 2$ interchange:
\bea
&& {1 \over k_1^2} \int_{\sigma_1}\int_{\sigma_1>{\sigma_1}'>{\sigma_1''}} 
\bigg[ {(\dot y_2(\sigma_2).\dot y_1({\sigma_1}''))
(\dot y_1(\sigma_1).\dot y_1({\sigma_1}')) \over 
4 \pi^2 |y_1(\sigma_1)-y_1({\sigma_1}')|^2} \ 
e^{ik_1.(y_2(\sigma_2)-y_1({\sigma_1}''))} \nonumber \\
&& \hspace{20 ex}  +{(\dot y_2(\sigma_2).\dot y_1({\sigma_1}'))
(\dot y_1(\sigma_1).\dot y_1({\sigma_1}'')) \over 
4 \pi^2 |y_1(\sigma_1)-y_1({\sigma_1}'')+\theta k_1|^2} \ 
e^{ik_1.(y_2(\sigma_2)-y_1({\sigma_1}'))} \nonumber \\
&& \hspace{20 ex} +{(\dot y_2(\sigma_2).\dot y_1(\sigma_1))
(\dot y_1({\sigma_1}').\dot y_1({\sigma_1}'')) \over 
4 \pi^2 |y_1({\sigma_1}')-y_1({\sigma_1}'')|^2} \  
e^{ik_1.(y_2(\sigma_2)-y_1(\sigma_1))} \bigg]
\label{anine}
\eea 
The $y_1$ structure of this term is much more complicated than that of
the first term in (\ref{aone}). So the result of applying the loop
laplacian on it is also more complicated. For example, let us consider
the first term in the above expression. If the loop laplacian acts on
$y_1({\sigma_1}'')e^{-ik_1.y_1({\sigma_1}'')}$, the result is simple
and, in fact, just reproduces that contribution of the second term on
the right hand side of the loop equation (\ref{foursix}) in which a
self-energy insertion is present on the contour $C_{1\tau}$ entirely
above the marked point $\tau$. A similar operation of the loop
laplacian on the other two terms in (\ref{anine}) reproduces the other
two contributions in which the self-energy insertion is either
entirely below the marked point $\tau$ or across it. On the other
hand, if the loop laplacian acts on the propagator $1/4 \pi^2
|y_1(\sigma_1)-y_1({\sigma_1}')|^2$, the result is a delta-function
type of contribution. Together with a similar contribution from the
last term in (\ref{anine}) (the middle term has no contribution of
this type since the delta-function does not click), this precisely
reproduces the entire contribution of the first term on the right hand
side of the loop equation in this order.

Let us now go to the next term, (\ref{atwo}). Here, the loop laplacian
may act on any of the four propagators, resulting in a delta-function.
This gives four different terms and these precisely reproduce that
contribution of the second term on the right hand side of the loop
equation (\ref{foursix}) in which a gauge boson is exchanged between
the two contours $C_{1\tau}$ and $C_{2s}$. For this it is essential to
remember that the loop on the right hand side, ${\rm Tr}(\hat
W_{C_{1\tau}}[y_{1\tau}] \hat W_{C_{2s}}[y_{2s}])$, involves the
hatted operators. As defined in (\ref{twotwelve}), these differ from
the unhatted ones in that the argument of the gauge field is shifted
by the starting point of the contour. For the contours $C_{1\tau}$ and
$C_{2s}$, the starting points are respectively
$y_{1\tau}(0)=y_1(\tau)$ and $y_{2s}(0)=y_2(s)$. The four terms
mentioned above correspond to the four differnt possibilities of the
two ends of the gauge field propagator landing above or below the
marked points on the two contours.

Thus, we see that the action of the loop laplacian on the second order
contribution to the $2$-point function reproduces all the terms
expected on the right hand side of the loop equation (\ref{foursix}).
There are also other terms produced in the process of applying the
loop laplacian on the contribution of individual diagrams to the
$2$-point function. We have seen above an explicit example of one such
term which, however, eventually disappeared because of a nontrivial
cancellation with another term. We expect that all other similar terms
(which are produced in the process of applying the loop laplacian on
the contributions of different diagrams to the $2$-point function, but
are not present on the right hand side of the loop equation) will
eventually disappear through similar cancellations, but we have not
attempted a complete verification of this.

\end{document}